\documentclass[reprint,12pt]{JASA}
\usepackage[utf8]{inputenc}
\usepackage[T1]{fontenc}
\usepackage{graphicx}
\usepackage{grffile}
\usepackage{wrapfig}
\usepackage{rotating}
\usepackage[normalem]{ulem}
\usepackage{amsmath}
\usepackage{textcomp}
\usepackage{amssymb}
\usepackage{capt-of}
\usepackage{hyperref}
\usepackage{graphicx}
\usepackage{cleveref}
\usepackage{siunitx}
\usepackage{float}
\usepackage{subfigure}
\usepackage{adjustbox}
\date{\today}
\title{}
\hypersetup{
 pdfauthor={thezuf},
 pdftitle={},
 pdfkeywords={},
 pdfsubject={},
 pdfcreator={Emacs 26.3 (Org mode 9.1.9)}, 
 pdflang={English}}
\begin{document}

\title{Ray tracing and finite element modeling of sound propagation in a compartment fire}

\author{Mustafa Z. Abbasi} \email{mustafa\_abbasi@utexas.edu}

\thanks{Corresponding author.}
\author{Preston S. Wilson}
\author{Ofodike A. Ezekoye}
\affiliation{Walker Department of Mechanical Engineering, The University of Texas at Austin, Austin, Texas }

\date{\today}

\begin{abstract}
A compartment fire (a fire in a room or building) creates temperature gradients and inhomogeneous time-varying temperature, density, and flow fields. This work compared experimental measurements of the room acoustic impulse/frequency response in a room with a fire to numerically modeled responses. The fire is modeled using Fire Dynamics Simulator (FDS). Acoustic modeling was performed using the temperature field computed by FDS. COMSOL™ Multiphysics was used for finite element acoustic modeling and Bellhop™ for ray-trace acoustics modeling. The results show that the fire causes wave-fronts to arrive earlier (due to the higher sound speed) and with more variation in the delay times (due to the sound speed perturbations). The frequency response shows that the modes are shifted up in frequency and high frequency (>2500 Hz) modes are significantly attenuated. Model results are compared with data and show good agreement in observed trends.
\end{abstract}

\maketitle

\section{Introduction}
\label{sec:orgef5e1fd}
\label{ch:chapter2_modelling_acoustics}

Fire changes the acoustics properties of a room by introducing temperature variations and time-dependent temperature and flow fields \cite{quintiere2006fundamentals}. Firefighters use acoustic alarms to locate and rescue downed firefighters on the fireground. This work aims to understand how sound propagation in a room changes due to a fire being introduced into the room. \citet{Abbasi2020_Change,abbasi2020Sound} showed that the measured acoustic impulse response of a room is significantly changed. Low-frequency modes increased in frequency, and higher-frequency modal structure was lost.  We hypothesize that the dominant mechanism for the measured changes in impulse response is the time-varying temperature field which leads to a time-varying sound speed field. To test this hypothesis, this work used numerical modeling, allowing the decoupling of some of these effects to isolate the dominant physical mechanism. Two types of sound propagation models (a ray model and a full-wave finite element model) were used with a sophisticated computational fluid dynamics (CFD)  fire model to simulate the effect of the fire on acoustic propagation. 

\section{Experimental Results}
\label{sec:org24021dc}
\begin{figure}[]
\centering
\includegraphics[keepaspectratio,width=1.0\linewidth,height=0.8\textheight]{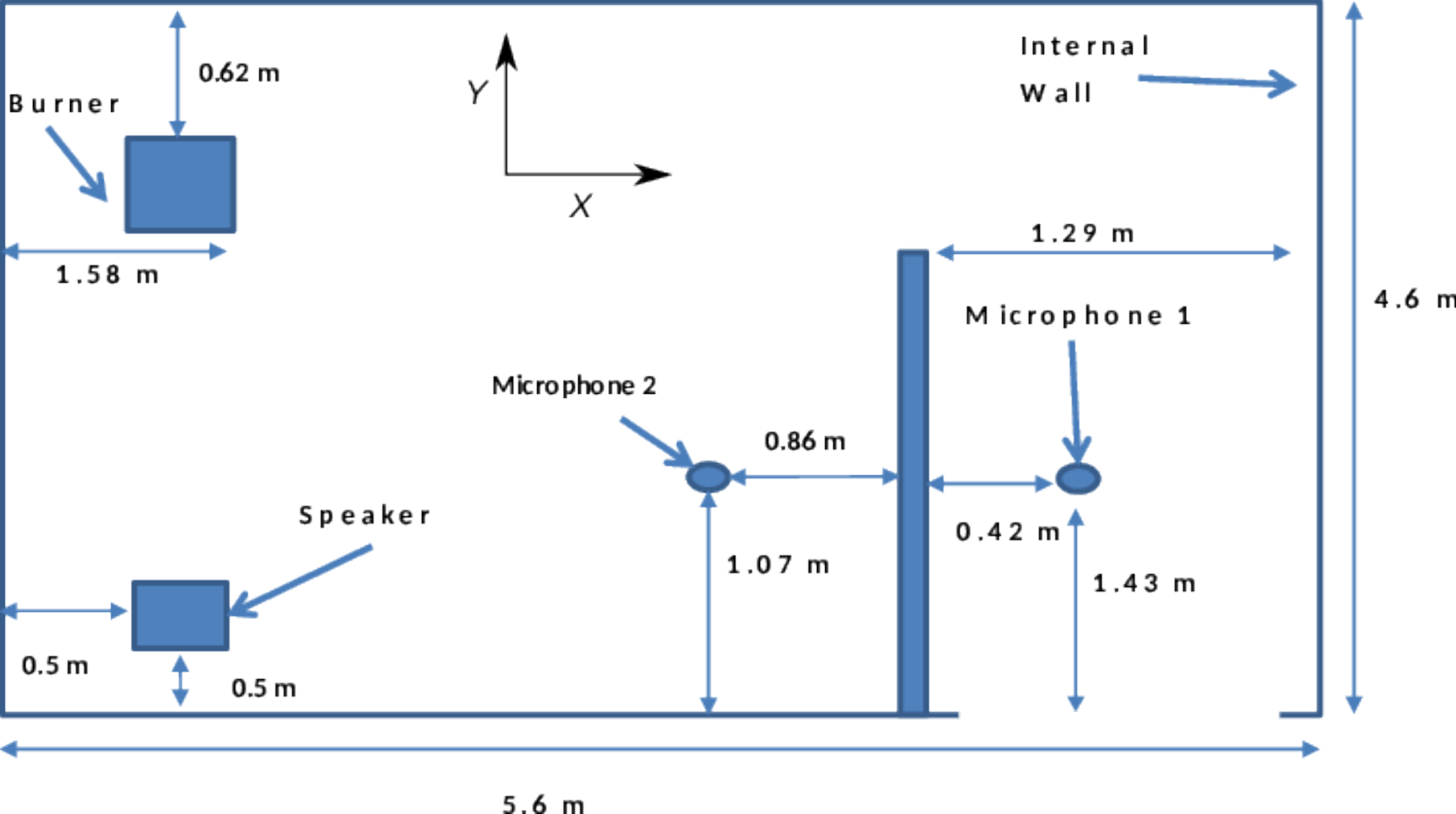}
\caption{\label{fig:exp1_schem}
(Color online) Top view of the burn compartment and equipment for experiment 1, described in \cite{Abbasi2020_Change,abbasi2020Sound}. Microphones were placed at height \(H\) = 0.56 m above the floor.}
\end{figure}

\begin{figure}[]
\centering
\includegraphics[keepaspectratio,width=1.0\linewidth,height=0.8\textheight]{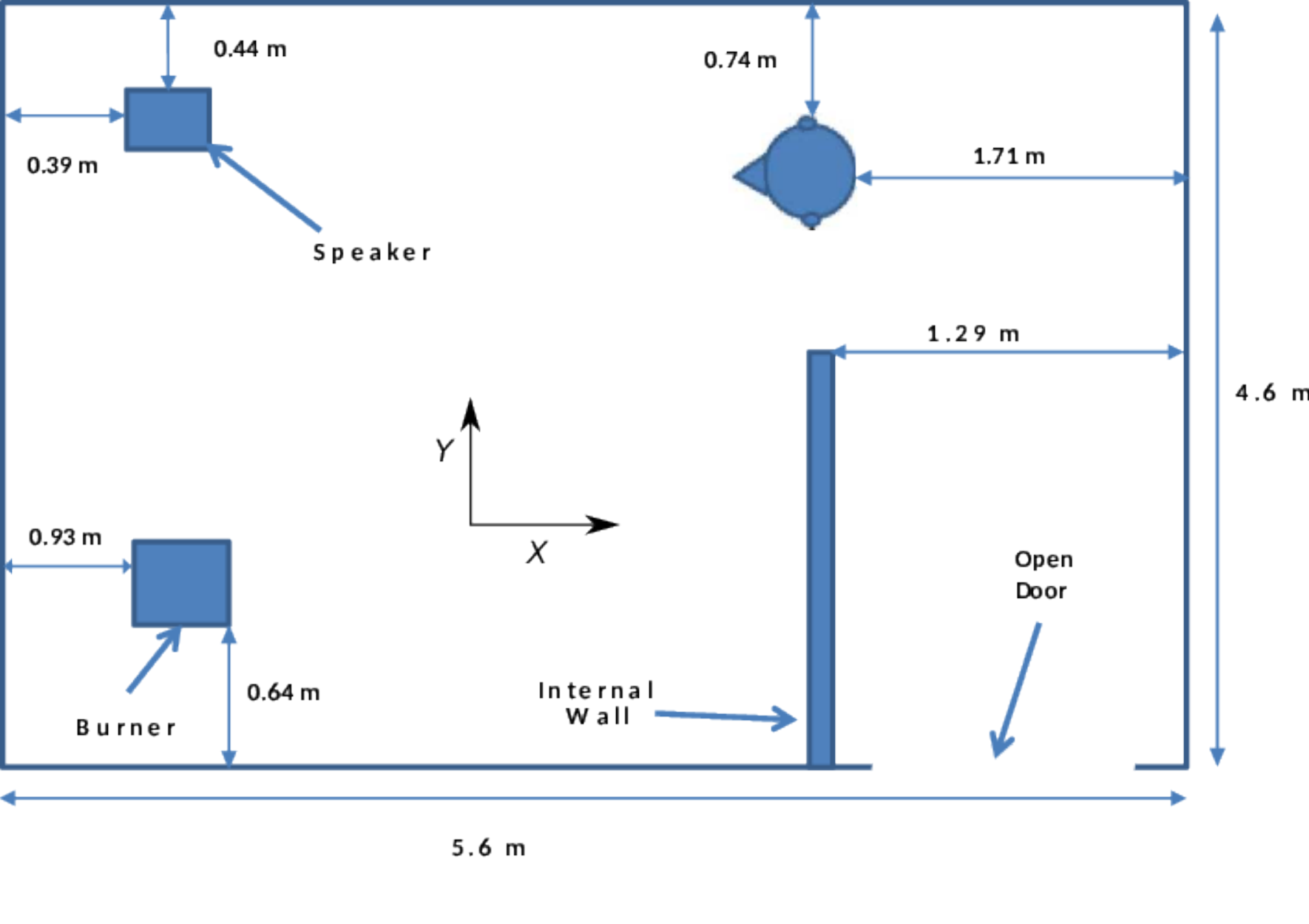}
\caption{\label{fig:exp3_schem}
(Color online) Top view of the burn compartment and equipment for experiment 3, described in \cite{abbasi2020Sound}. The glass manikin head is in the upper right corner of the figure, facing in the \ensuremath{X}-negative direction (toward the speaker).}
\end{figure}

The numerical models developed in this article are compared with experimentally measured impulse responses previously shown in \citet{Abbasi2020_Change,abbasi2020Sound}. Experiments 1 and 3 (according to the nomenclature introduced in \citet{abbasi2020Sound} are modeled).  \Cref{fig:exp1_schem} and \Cref{fig:exp3_schem} shows the schematic diagrams for those two experiments. For experiment 1, the impulse/frequency response measured by microphone 2 is used for comparison. For experiment 3, the impulse/frequency response measured by the `left ear' microphone is used for comparison.

\section{Fire Modeling}
\label{sec:orgfd5bc75}
\label{sec:fds_model}

A fire is a reaction between fuel and oxygen releasing heat and chemical byproducts. The experimental compartment fires of \citet{Abbasi2020_Change} were modeled using the open-source finite difference CFD model Fire Dynamics Simulator version 6 (FDS). The large-eddy turbulence model in FDS \cite{mcgrattan2013fire} was used to simulate the three-dimensional temperature field created by the fire. FDS was used to compute the spatial and temporal evolution of the temperature field (two-dimensional and three-dimensional) which were used as input to the acoustic models. In \Cref{sec:comsol_scattering}, the model is used to simulate a three-dimensional temperature field created by a fire in an open environment. The fire is based on the one used in the experimental measurements described in \citet{Abbasi2020_Change}, without any compartment effects. In \Cref{sec:2d_fire_comsol} the experimental compartment fires shown in \citet{Abbasi2020_Change} are simulated. A two-dimensional slice of the temperature field was taken for each time step to capture the direct path between the acoustic source and receiver. The fire was modeled as a planar surface with a specified heat release rate (HRR). The walls, floor, and ceiling were modeled as 16-cm-thick gypsum.

\section{Finite Element Acoustic Model}
\label{sec:org5437aea}
\label{sec:comsol_model_overview}

\begin{figure}[]
\centering
\includegraphics[angle=0,width=1.0\linewidth, keepaspectratio,,height=0.8\textheight]{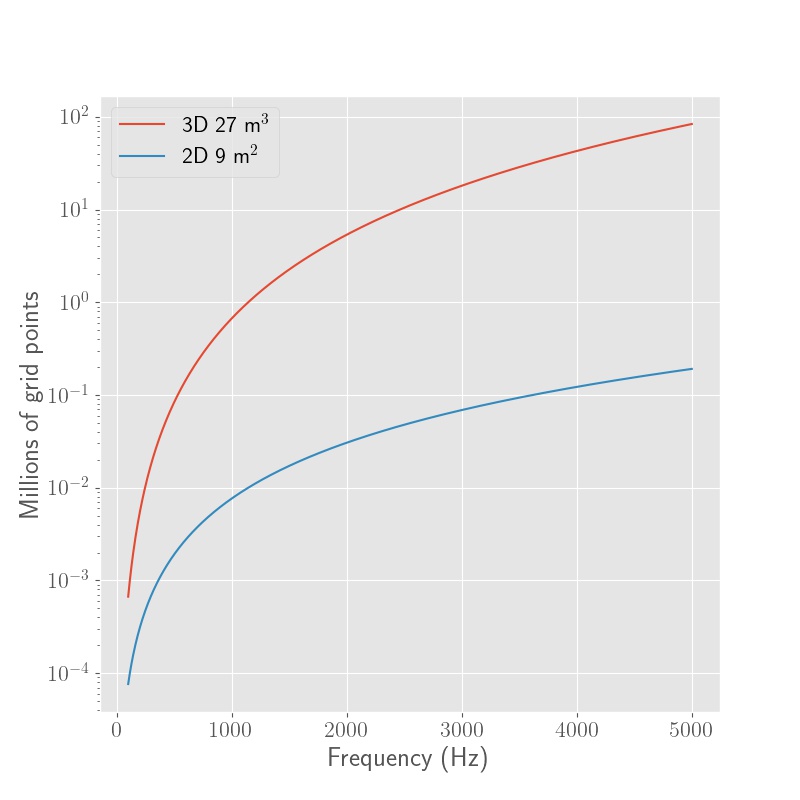}
\caption{\label{fig:model_n_point_comsol}
(Color online) Number of discrete cells required as a function of frequency for finite element modeling in two and three dimensions for a \SI{3}{\meter\cubed} room.}
\end{figure}

The time-domain wave equation, and its frequency-domain analog, the Helmholtz equation, govern linear sound propagation. The Helmholtz equation can be solved numerically using the finite element method (FEM). To construct the model, the geometry of interest is discretized into small elements on which the discretized form of the Helmholtz equation is solved. For accurate solution, the geometry must be discretized with  maximum element size < \(\frac{ \lambda }{ 10 }\) where \(\lambda\) is the wavelength \cite[p. 554]{multiphysics_manual_4p3}. One consequence of this is that for frequencies of interest for the PASS problem (500~Hz to 5000~Hz), this can require hundreds of millions of grid points as shown in \Cref{fig:model_n_point_comsol}. Therefore, three-dimensional modeling of the compartment fire acoustics problem was found to be impractical. This work will limit the acoustic modeling to two-dimensional slices. It is important to note that our purpose in this modeling was understanding the acoustics of the room fire and to gain insight into the experimental results, not as a design or auralization tool.  This is analogous to calculating transmission loss along a radial in underwater acoustics, though the authors acknowledge that out-of-plane effects will be significant in this geometry, while they can be negligible in some underwater acoustics problems. Because of the two-dimensional nature of this modeling, out-of-plane acoustic paths are not present in the model and therefore comparisons between model and measurement may exhibit differences because of this approximation.  It is useful nonetheless to test the degree to which this expedient approximation remains valid. Hence in this work, we focus on trends in the results rather than a quantitative comparison. In the limiting case of a long narrow hallway, the model and data would be more directly comparable. Also, we limit the model to frequency-independent losses, to isolate path changes due to temperature variations as the fundamental difference between iso-velocity and compartment fire acoustic propagation. We believe despite these limitations, this model is a step forward in modeling compartment fire acoustics.

The commercial finite element package COMSOL Multiphysics (version 4.2) was used in this work. The finite element model was computed on the temperature distribution calculated by FDS. The flow field is ignored, thus isolating the effect of temperature variations. This assumption is valid since the velocities of the flows are insignificant compared to the lowest sound speed in the model.

\subsection{Three-dimensional finite element model: scattering from the bare flame.}
\label{sec:org113e205}
\label{sec:comsol_scattering}

The fire creates chaotic temperature and flow fields that could scatter sound. Previous work by \cite{abbasi2013development} showed that a flame scattered acoustic energy and thereby impacted the accuracy of an acoustic range finder operated through the flame. This section describes the results of a three-dimensional finite element acoustics model coupled with a three-dimensional finite-difference CFD fire model (COMSOL and FDS) to understand the effect of the flame on the sound propagating through the flame.

\begin{figure}[]
\centering
\includegraphics[angle=0,width=1.0\linewidth, keepaspectratio,,height=0.8\textheight]{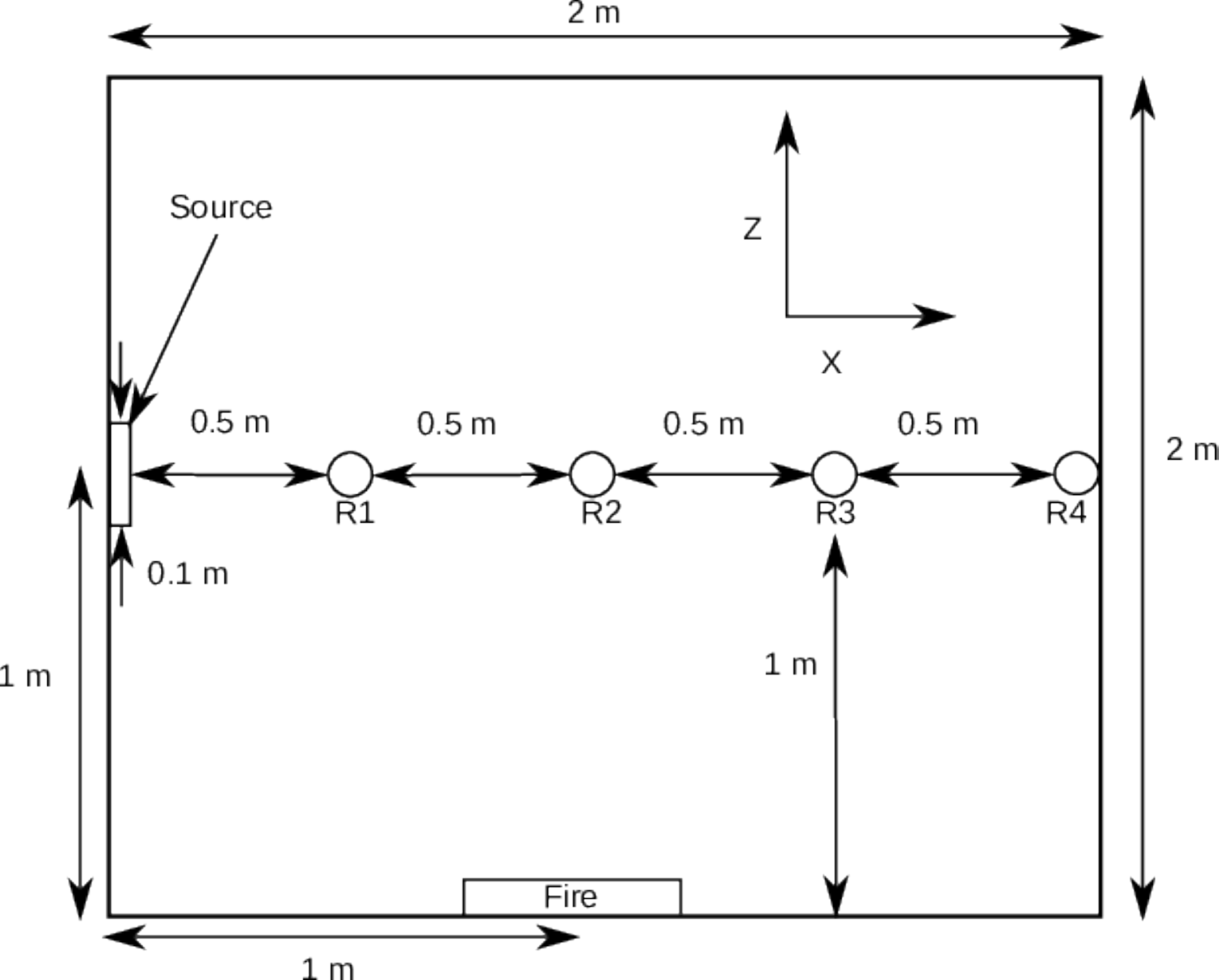}
\caption{\label{fig:3d_fire_scattering_1}
(Color online) Diagram of the three-dimensional sound propagation simulation discussed in \Cref{sec:comsol_scattering}. A fire is placed in a free field, computed in a cubic computational domain \SI{2}{\meter} per side, with four receivers placed in front of a source.}
\end{figure}

A COMSOL model was constructed to compute the acoustic pressure \(P(f,T)\) as a function of time \(T\) and frequency \(f\). A measurement of the change in acoustic level due to the fire is given by \(\Delta RL(f,T) =  10 \log_{10} (\frac{P(f,T)}{P(f,T=0)})^2\) which was computed in post-processing. \Cref{fig:3d_fire_scattering_1} shows a diagram of the domain. The domain consists of a \SI{2}{\meter}~x~\SI{2}{\meter}~x~\SI{2}{\meter} space. Coordinate positions for the receiver, source and fire are shown in \Cref{tab:comsol_poses}. The source is a \SI{0.1}{\meter} x \SI{0.1}{\meter} plane with a constant source amplitude of 1 Pa. 

\begin{table}[]
\caption{\label{tab:comsol_poses}
(Color online) Table of receiver, source and fire positions for the three-dimensional COMSOL}
\begin{tabular}{|c|c|c|c|}
\hline
 & \(X\) (m) & \(Y\) (m) & \(Z\) (m)\\
\hline
Burner & 0.0 & 0 & 0\\
Source & \(-1.0\) & 0 & 1\\
Receiver R1 & \(-0.5\) & 0 & 1\\
Receiver R2 & 0.0 & 0 & 1\\
Receiver R3 & 0.5 & 0 & 1\\
Receiver R4 & 1.0 & 0 & 1\\
\hline
\end{tabular}
\end{table}

The simulated fire matches the properties of the burner used in the experimental measurements described in \cite{Abbasi2020_Change}. It has a square profile \SI{0.3}{\meter} x \SI{.3}{\meter}. COMSOL was run the frequency domain acoustics mode, a frequency sweep from 200~Hz to 900~Hz, with \(\delta f\)~=~1~Hz. The model was recomputed every 0.1~s for 2~s. The fire was modeled in FDS and the three-dimensional temperature field computed is input to COMSOL at each computational time step. FDS was run with 3.125~cm grid resolution. The COMSOL mesh was recomputed every 100 Hz with maximum element size \(\frac{\lambda}{10}\), and \(c_0 =\) \SI{500}{\meter\per\second}. Plane-wave radiation conditions were applied in COMSOL to the boundaries of the geometry, and in FDS the boundaries were assigned open. This was done to approximate a flame in a free field, with a source on one side, and receivers in front of the source.

\begin{figure}[]
\centering
\includegraphics[angle=0,width=1.0\linewidth, keepaspectratio,,height=0.8\textheight]{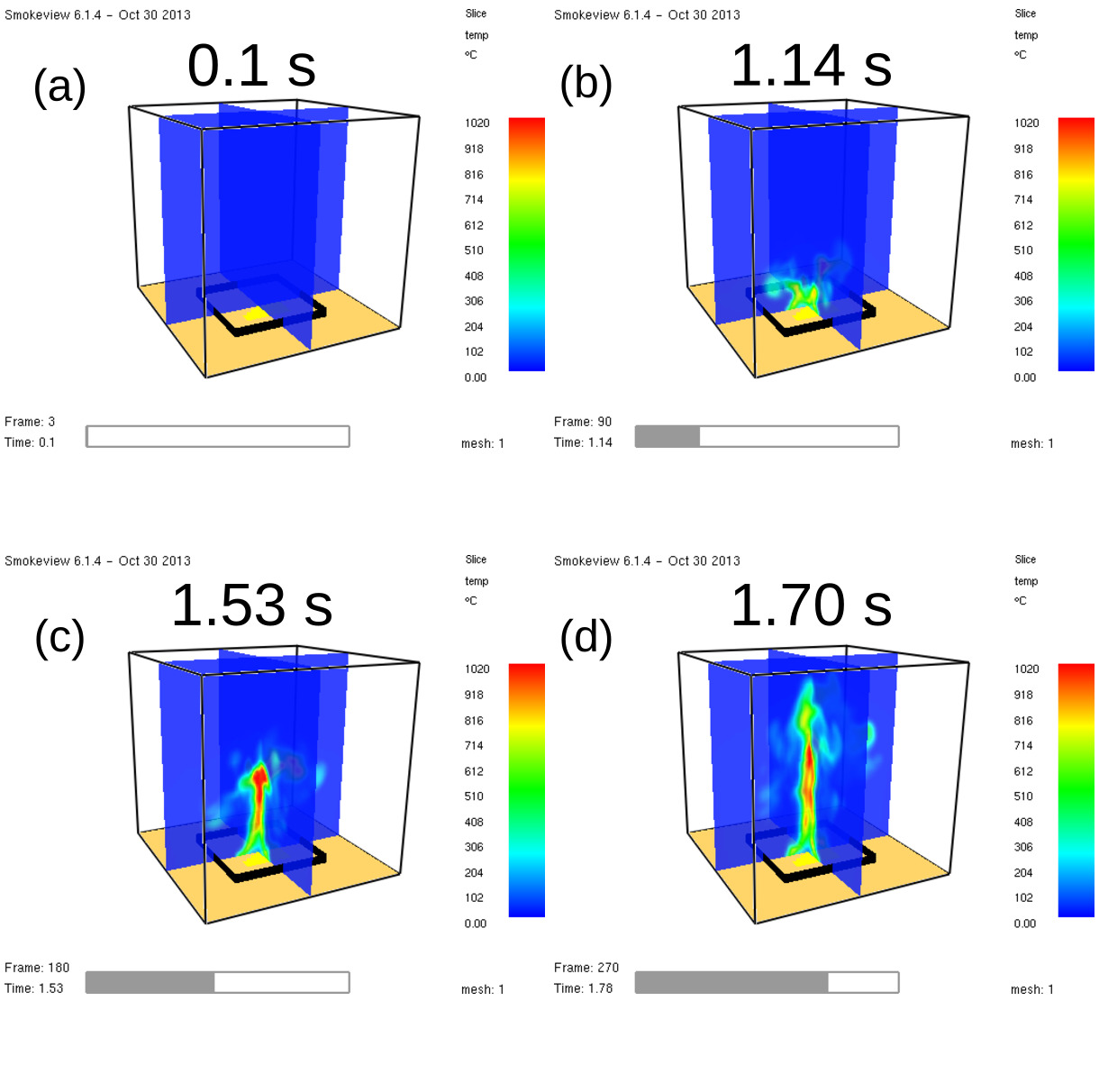}
\caption{\label{fig:3d_fire_scattering_2}
(Color online) The temperature field created by the fire used for the three-dimensional fire/acoustic model discussed in \Cref{sec:comsol_scattering}. The field 0.1~s after the start of the simulation, at the time of ignition is shown in (a). The field is shown at subsequent times in (b), (c), and (d).}
\end{figure}

\begin{figure}[]
\centering
\includegraphics[angle=0,width=1.0\linewidth, keepaspectratio,,height=0.8\textheight]{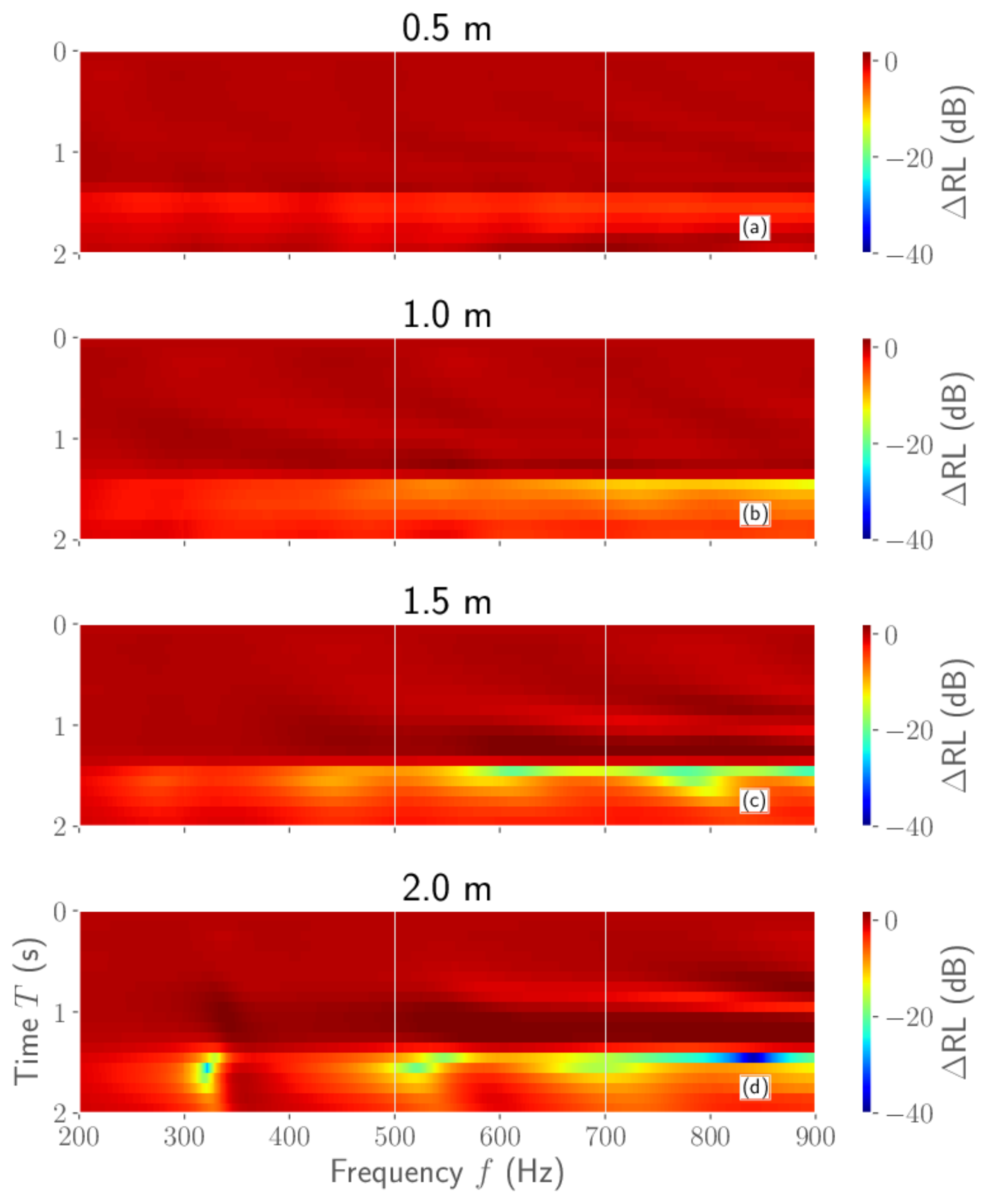}
\caption{\label{fig:COMSOL_results_plot_3d_fds}
(Color online) The change in acoustic pressure, \(\Delta RL(f,T)\),  is shown at four different receiver positions in front of a source over time as a flame develops in the environment.}
\end{figure}

\begin{figure}[]
\centering
\includegraphics[angle=0,width=1.0\linewidth, keepaspectratio,,height=0.8\textheight]{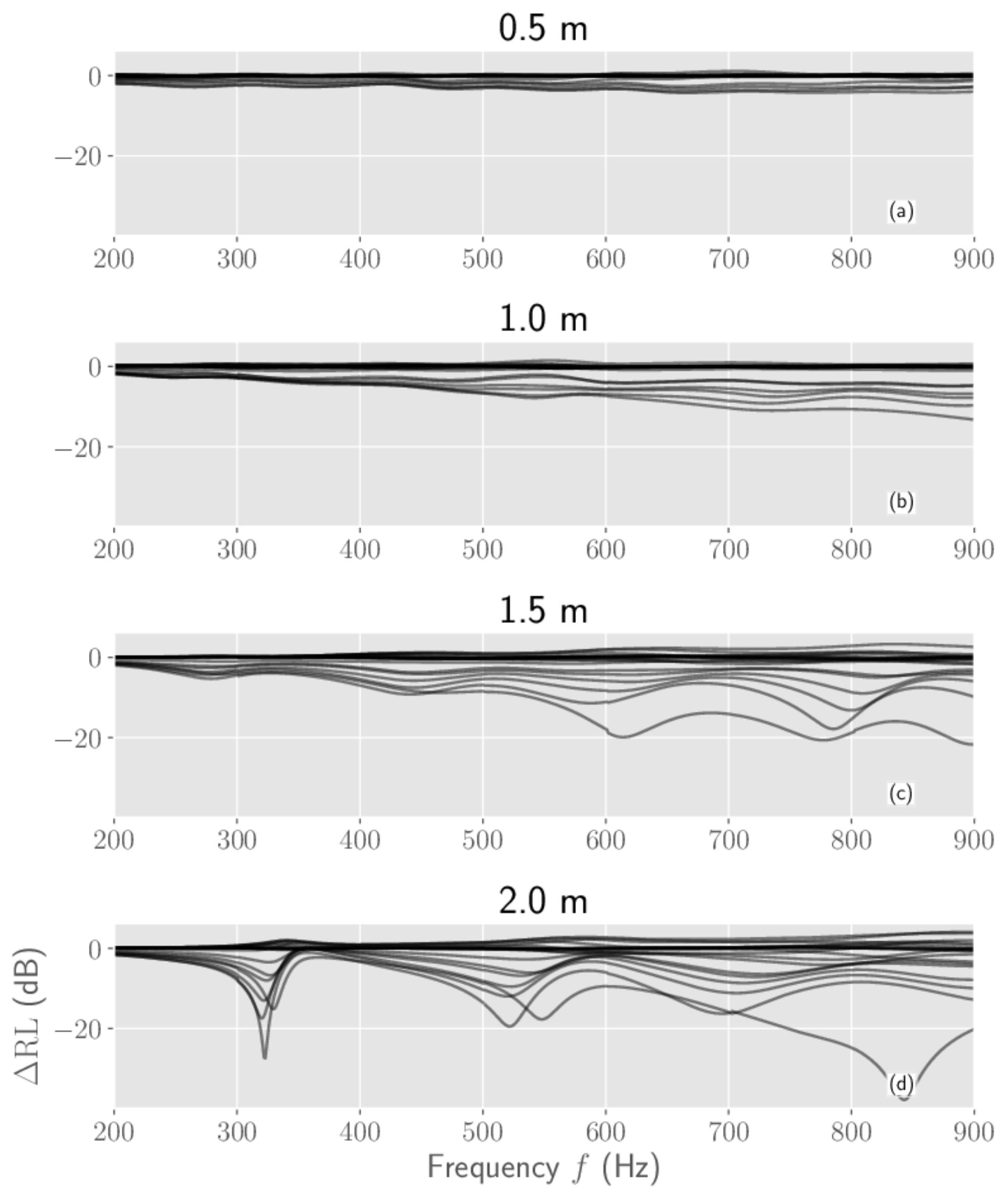}
\caption{\label{fig:COMSOL_results_plot_3d_fds_max_change_line}
(Color online)   The change in acoustic pressure, \(\Delta RL(f,T)\),  is shown at four different receiver positions in front of a source over time. Each subfigure shows the \(\Delta RL(f,T)\) for every time step for 2 s with \(\delta t\) = 0.1 s}
\end{figure}

\Cref{fig:3d_fire_scattering_2} shows the temperature field at four example times over the course of the model run and  \Cref{fig:COMSOL_results_plot_3d_fds} shows \(\Delta RL(f,T)\) as a function of time and frequency. At \ensuremath{T}~=~0~s no fire is present, and that is considered the baseline condition. As the fire develops, there is a change in the received acoustic pressure spectrum at all of the receivers. The receiver closest to the source is impacted least, and low frequencies are impacted the least. The greatest \(\Delta RL(f,T)\) occurs when the flame impinges on the horizontal plane containing the receivers. \Cref{fig:COMSOL_results_plot_3d_fds_max_change_line} shows the  fire is acting as a notch filter.

\subsection{Propagation in a two-dimensional slice through a compartment fire}
\label{sec:org9e0d08b}
\label{sec:2d_fire_comsol}

\begin{figure}[]
\centering
\includegraphics[angle=0,width=1.0\linewidth, keepaspectratio,,height=0.8\textheight]{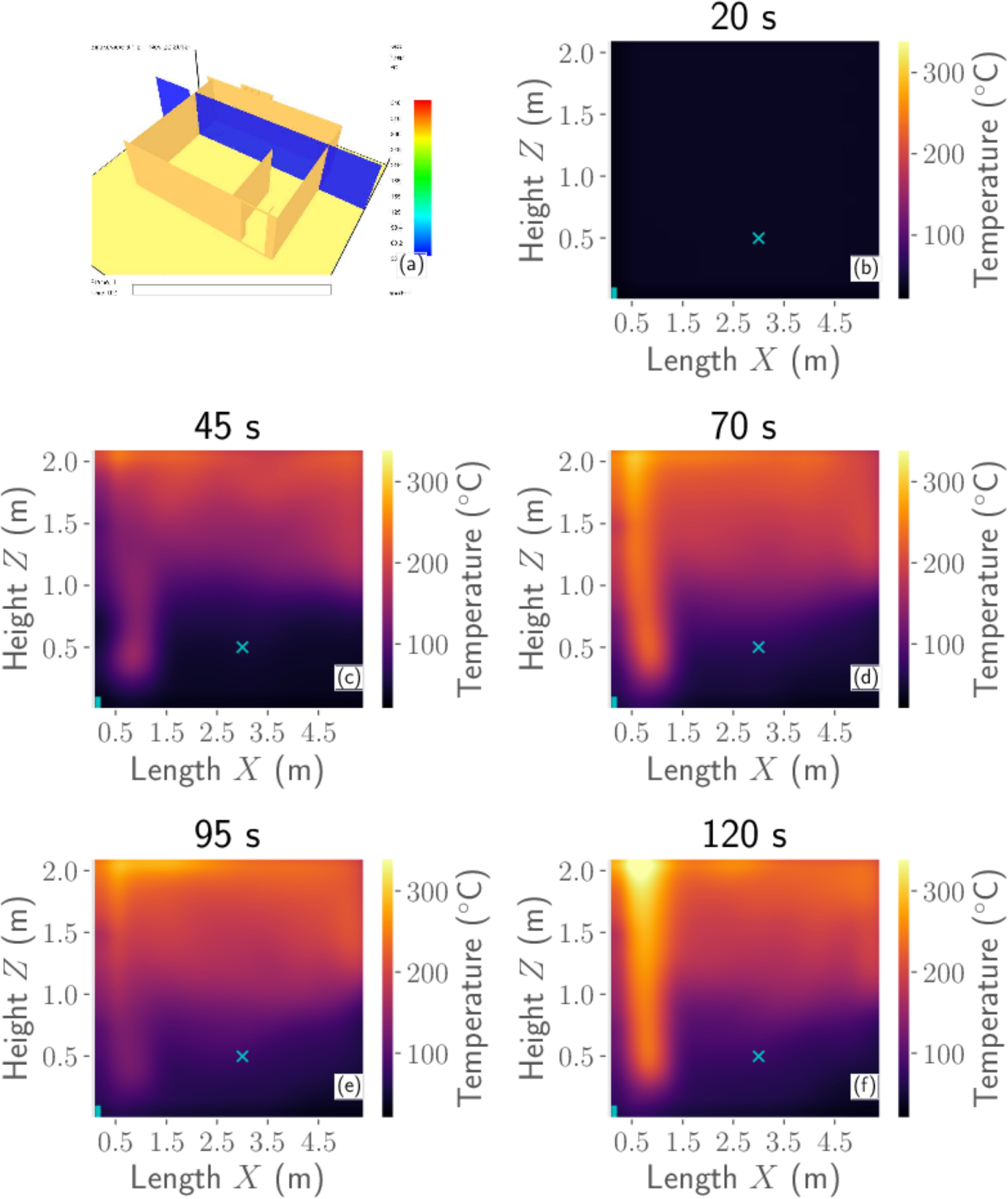}
\caption{\label{fig:fds_grid_plot}
(Color online) Results of CFD fire model showing the temperature in a two-dimensional plane as a function of time. The location of the slice is shown in (a). The cyan area (lower left on the subfigures) marks the source and the cyan `x' marks the receiver position.}
\end{figure}

The experimental room for compartment fire experiments described in \cite{Abbasi2020_Change} was modeled using FDS. The vertical two-dimensional slice chosen was 0.5 m from the burner (shown as the blue slice in the top left subfigure of \Cref{fig:fds_grid_plot}). The simulated thermal fields are shown in \Cref{fig:fds_grid_plot} at various times. The temperature increases and then stabilizes as the system reaches a steady-state. The FDS model included a 413 kW fire, an open door in the hallway, and gypsum walls. The model is discretized using 128 x 128 x 32 grid points, resulting in 7.5~cm x 7.5~cm x 10~cm (\(X\), \(Y\), \(Z\)) grid resolution. The fire was ignited at \(T\)~=~20~s and allowed to run till \(T\)~=~120 s.

The finite element acoustics model used a two-dimensional geometry (\(X=2.1\)~m,  \(Z = 5.1\)~m) with rigid boundary conditions. The COMSOL frequency domain module conducted a parametric sweep over frequency, with the \(\delta f\)~=~1~Hz. Re-meshing was a substantial run time expense; therefore, the mesh was regenerated every 500~Hz between 1~Hz and 2000~Hz and every 100~Hz between 2000~Hz and 4000~Hz. The highest frequency in the interval was used to compute the maximum element size. The acoustic source was placed in the lower-left corner (\(X\) = 0--0.1 m, \(Z\) = 0--0.1 m) of the vertical slice shown in \Cref{fig:fds_grid_plot}. The source was a 100 cm\(^2\) area with a constant 1 Pa source level. The point receiver was placed at (\(X = 3, Z = 0.5\)), modeling a crawling firefighter. The temperature slice was output from FDS every 1~s and input as the air temperature in the COMSOL model. 

\begin{figure}[]
\centering
\includegraphics[angle=0,width=1.0\linewidth, keepaspectratio,,height=0.8\textheight]{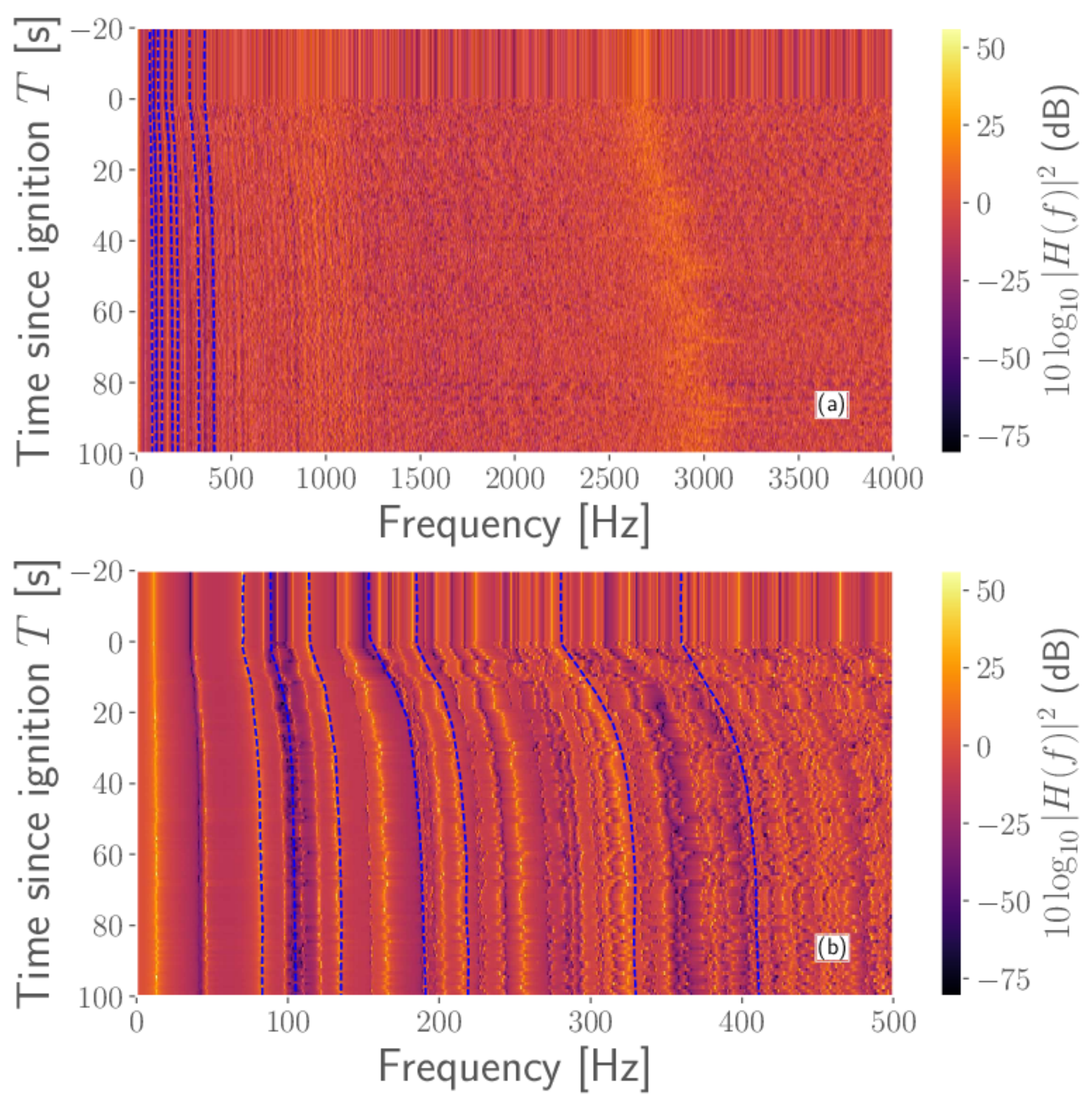}
\caption{\label{fig:comsol_0_4k}
(Color online) Finite element modeled acoustic frequency response for a source/receiver pair placed in a two-dimensional slice through a compartment fire. Ignition is at \(T\) = 0 s. Subfigure (a) shows the full band from 1~Hz to 4000~Hz, and Subfigure (b) shows the band from 1~Hz to 500~Hz. The blue lines track the frequency of select modes.}
\end{figure}

\Cref{fig:comsol_0_4k} shows the frequency response of the system computed using this model. Before ignition, the response is stationary. Modal peaks are visible and consistent. Introducing the fire into the compartment results in a time-varying unsteady frequency response. Low-frequency modes increase in frequency, and higher-frequency modal structure disappears. The model results show characteristics similar to the experimental results in \cite{Abbasi2020_Change}; low-frequency modes increase in frequency, high-frequency modes are less prominent, and the frequency response is highly time-varying after ignition. The dashed lines indicate modes whose frequencies were manually tracked over time.

\section{Two-Dimensional Ray Model in a Compartment Fire}
\label{sec:orgb7b0611}
\label{sec:ray_methods_overview}

\begin{figure*}
\centering
\includegraphics[angle=0,width=1.0\linewidth, keepaspectratio,,height=0.8\textheight]{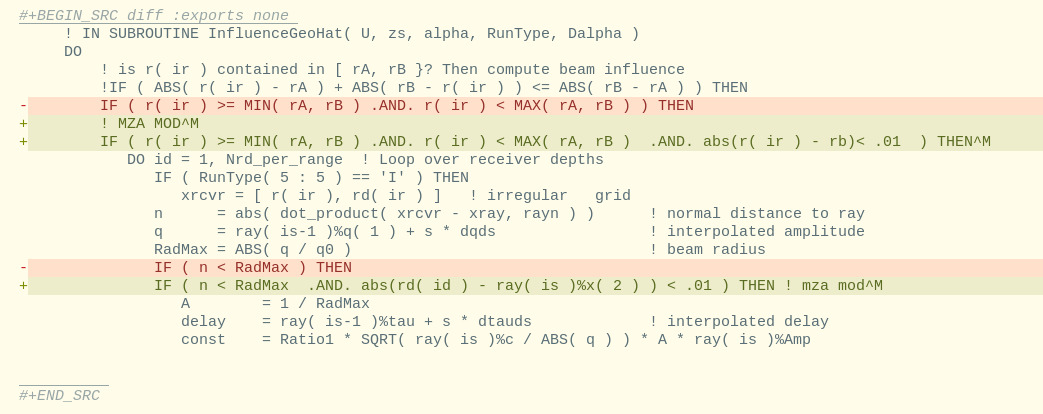}
\caption{\label{fig:bellhop_changes}
(Color online) A line by line difference between the original bellhop source code (bellhop.f90 source file) (in red) and the modifications made that code (in green).}
\end{figure*}

Ray theory is derived from the wave equation and considers acoustic paths (or rays) that follow Snell's law \cite{blackstock2000fundamentals,jensen2011computational}. Because of its geometric nature, ray theory is computationally efficient compared to full-wave methods.  Ray models used for room acoustics have traditionally been limited to iso-velocity environments, primarily because typical room acoustics applications do not include temperature/sound speed variations \cite{savioja2015overview}. A room with a fire has significant sound speed variation and therefore we cannot use traditional room acoustics software. Ray models used in underwater acoustics typically take sound speed variations into account and have been shown to provide an excellent comparison with measured data \cite{urick1983principles}. Therefore, we used an existing open-source underwater acoustics ray trace software, BELLHOP \cite{porter2007bellhop}. BELLHOP uses a predictor-corrector scheme to model ray paths. BELLHOP can output ray paths, eigenrays, and transmission loss at receiver locations. The version of BELLHOP used in this work is range-dependent and two-dimensional \cite{porter2011bellhop}. The BELLHOP code was modified (shown in \Cref{fig:bellhop_changes}) to add a constraint to ensure eigenrays always intersected with the receiver location within \textpm{}\SI{1}{\cm}. The model is limited to specular reflections only.

Ray paths and delay times at the receiver were computed by providing BELLHOP with a sound speed field from the FDS model described in \Cref{sec:fds_model} at each 1-second interval. The ray trace field was a rectangular region representing a vertical slice in the burn structure. A reflection coefficient of 0.4 dB was applied to all boundaries. The rectangular geometry and source/receiver position were adjusted to match the experiment being modeled.

Ten thousand rays were launched from an omnidirectional source, with launch angles equally spaced between \SI{0}{\degree} and \SI{360}{\degree}. For each ray trace, approximately 2000 arrivals are recorded. At each time step, the eigenray delays and amplitudes were computed. Let \(A_{n}\) and \(D_{n}\) be the amplitude and delay for arrival \(n\). The frequency response \(H(f)\) was computed using \Cref{eq:ray_freq_resp}, 
\begin{equation} 
\label{eq:ray_freq_resp}
H(f) = \sum_{n}^{N} A_{n} e^{-i 2 \pi f D_{n}},
\end{equation}
where \(f\) is the frequency in~Hz. The amplitude of each arrival is frequency-independent. The frequency response is the coherent sum of all arrivals. By assuming frequency-independent amplitude, we isolate the sound speed perturbations as the dominant mechanism for any changes in the impulse and frequency response.

\begin{figure}[]
\centering
\includegraphics[angle=0,width=1.0\linewidth, keepaspectratio,,height=0.8\textheight]{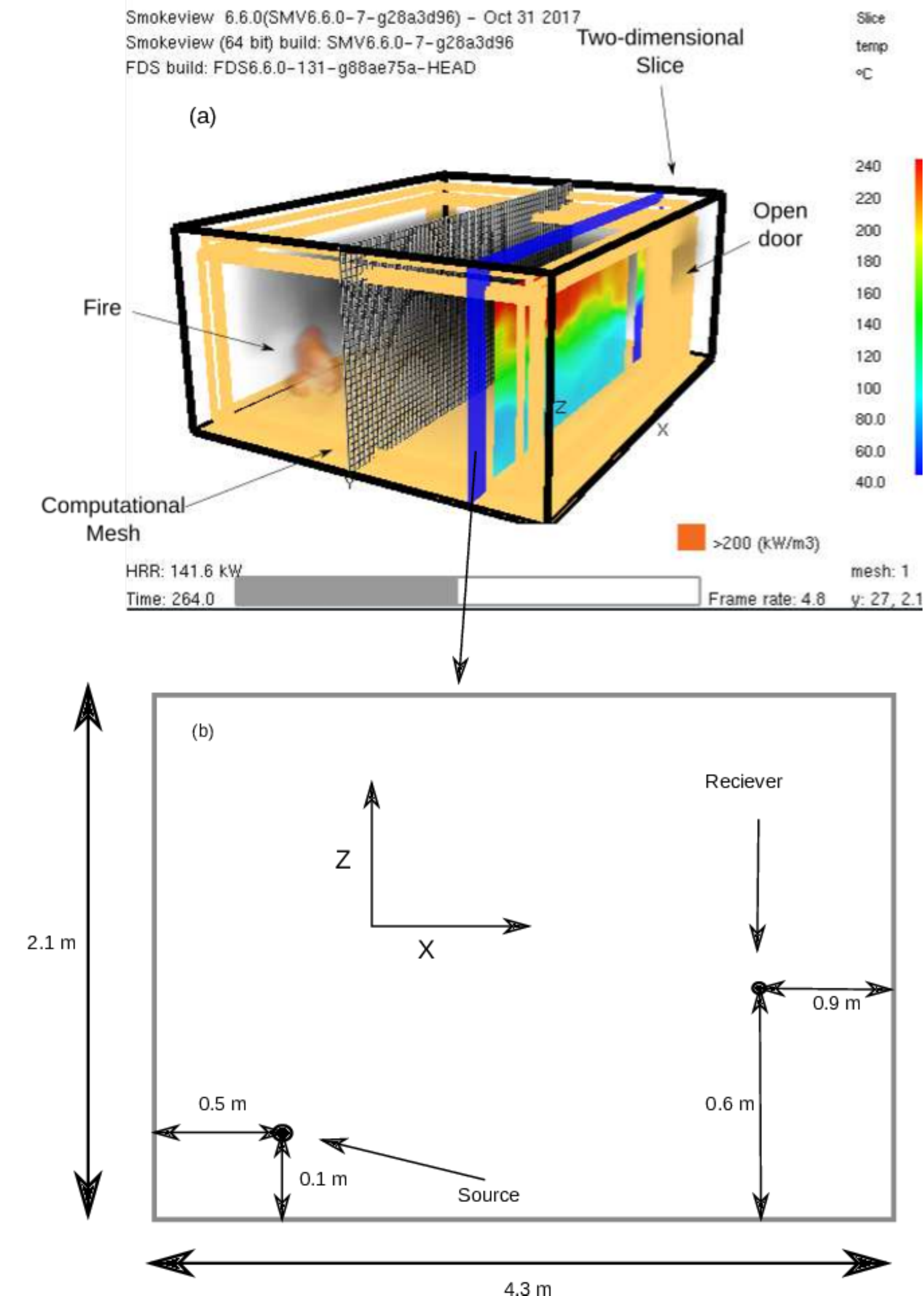}
\caption{\label{fig:FDS_model_schem_exp1}
(Color online) Visualization of experiment 1 modeled in FDS is shown in (a). The temperature at the  \(Y\)~=~0.5 m plane is shown at \(T\)~=~264.0 s. Ray model schematic for experiment 1 is shown in (b), showing source and receiver positions.}
\end{figure}

\begin{figure}[]
\centering
\includegraphics[angle=0,width=1.0\linewidth, keepaspectratio,,height=0.8\textheight]{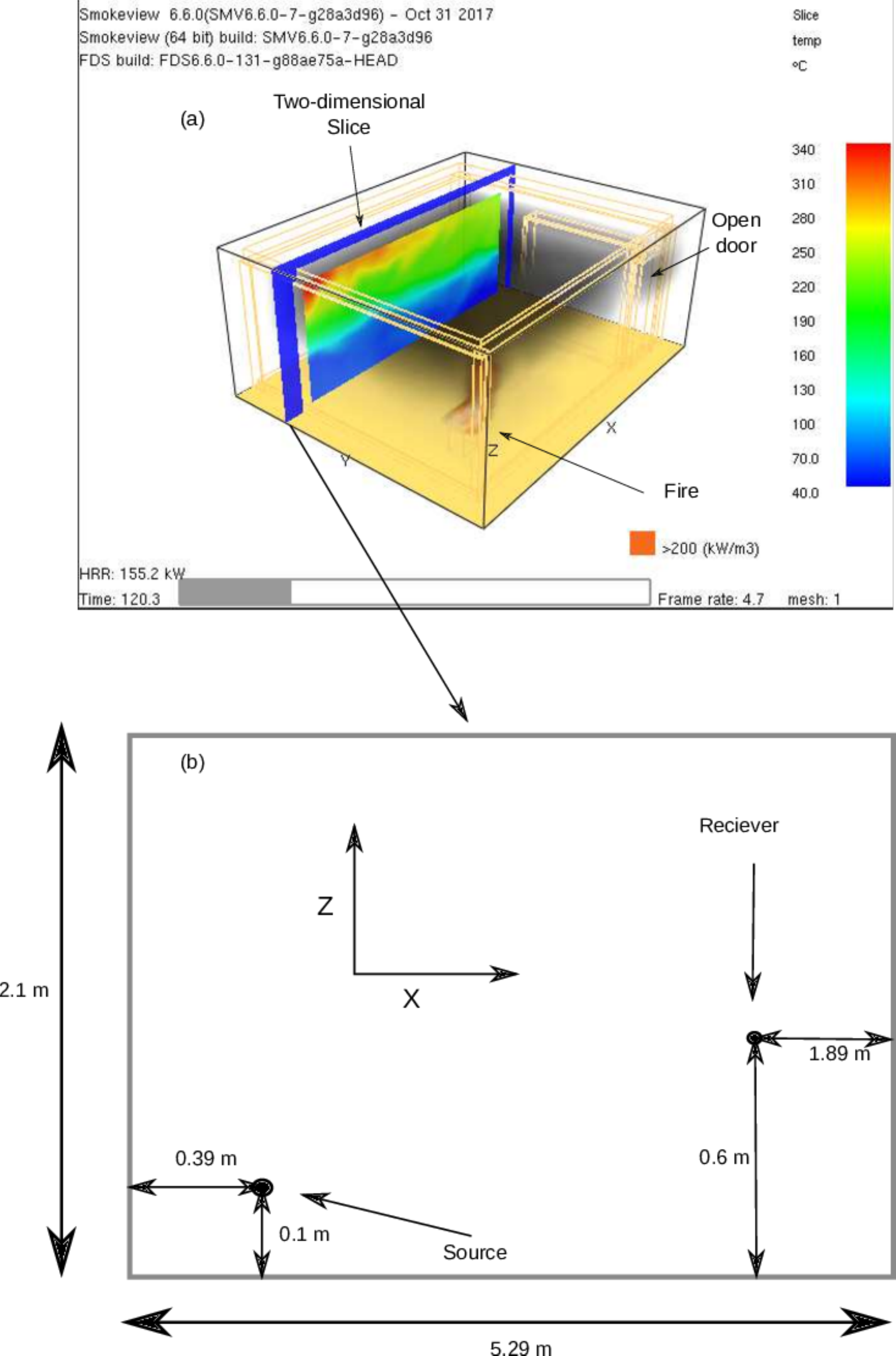}
\caption{\label{fig:FDS_model_schem_exp3}
(Color online) Visualization of experiment 3 modeled in FDS is shown in (a). The temperature at the  \(Y\)~=~3.5 m plane is shown at \(T\)~=~120.3 s. Ray model schematic for experiment 3 is shown in (b), showing source and receiver positions.}
\end{figure}

Experiment 1 and Experiment 3 were modeled. FDS used a mesh resolution of 0.10~m x 0.10~m x 0.09~m. The heat release rate (HRR) was set to  150~kW to match the experiments. The fire, acoustics source and acoustic receiver were positioned based on \Cref{fig:exp1_schem} and \Cref{fig:exp3_schem}. Visualization of the FDS models used to model experiment 1 (\Cref{fig:FDS_model_schem_exp1}(a)) and experiment 3 (\Cref{fig:FDS_model_schem_exp3}(a)) are shown. The flame is in one corner of the room. The visualization shows the geometry of the compartment, the two-dimensional plane of interest (with temperature marked by color), the vertical computational grid, the \(X\), \(Y\), and \(Z\) axes, and the smoke exiting the compartment.  Between the two experiments, the position of the fire and the acoustic source/receiver changes. \Cref{fig:FDS_model_schem_exp1}(b) and \Cref{fig:FDS_model_schem_exp3}(b) show source/receiver positions in the \ensuremath{X}-\(Z\) plane for experiment 1 and 3 respectively.

\begin{figure}[]
\centering
\includegraphics[angle=0,width=1.0\linewidth, keepaspectratio,,height=0.8\textheight]{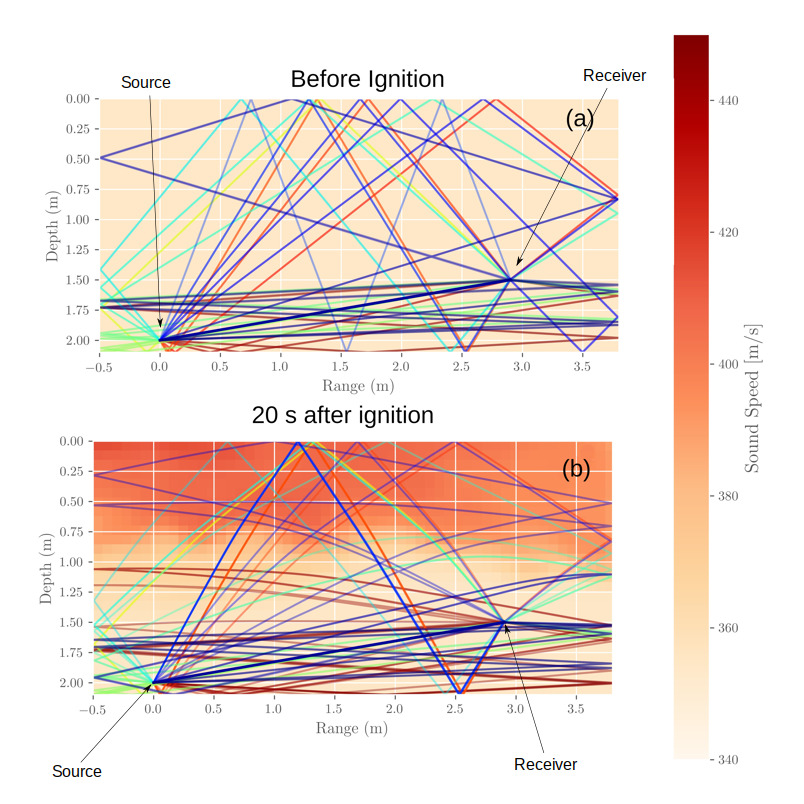}
\caption{\label{fig:ray_model_exp2_time_2}
(Color online) Ray paths modeling experiment 1 before ignition (a) and 20 s after ignition (b). This is visualization of the ray paths, limited to three boundary interactions.}
\end{figure}

\Cref{fig:ray_model_exp2_time_2} shows the instantaneous ray trace at two times (before ignition, and 20 s after ignition) modeling experiment 1. The fire changes the environment, resulting in rays launched at the same angle taking different paths. The additional floor interactions would result in reflection loss each time, increasing transmission loss over distance from the source. The change in ray paths could also change the perceived location of the PASS alarm.

\begin{figure}[]
\centering
\includegraphics[angle=0,width=1.0\linewidth, keepaspectratio,,height=0.8\textheight]{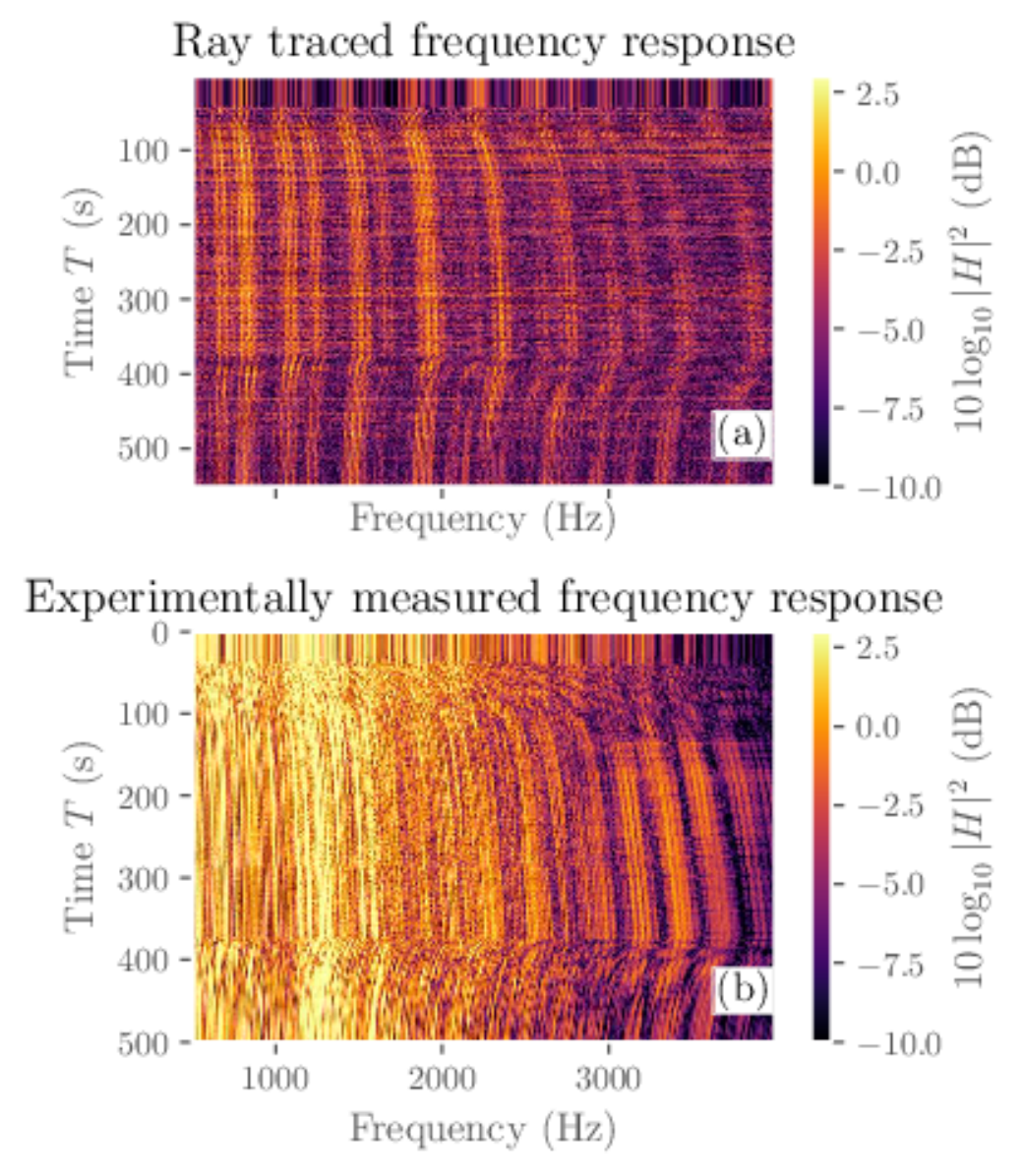}
\caption{\label{fig:ray_model_data_freq_resp_comparision_exp_2}
(Color online)  Modeled (a) and measured (b) frequency response for experiment 1.}
\end{figure}

\begin{figure}[]
\centering
\includegraphics[angle=0,width=1.0\linewidth, keepaspectratio,,height=0.8\textheight]{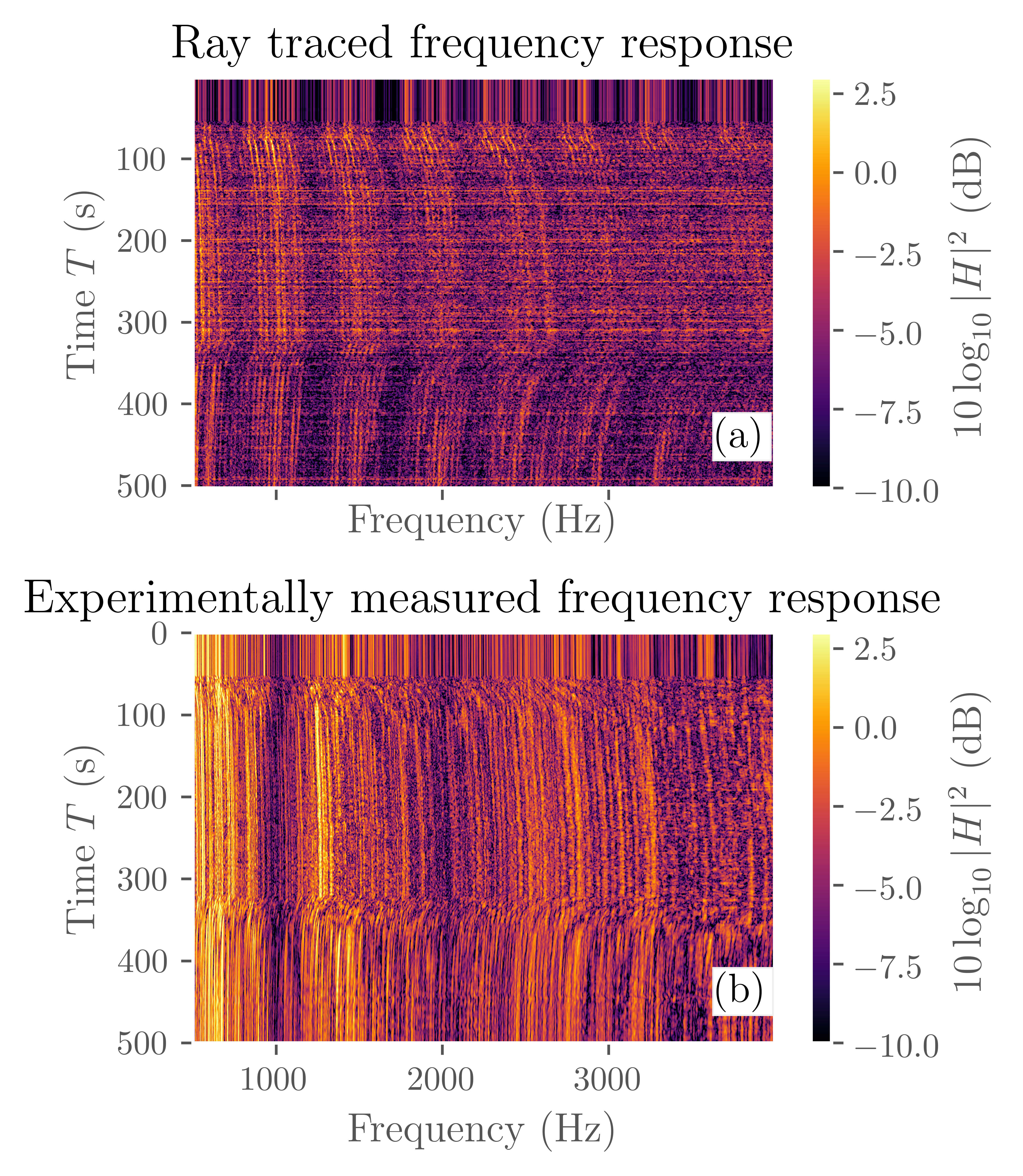}
\caption{\label{fig:ray_model_data_freq_resp_comparision_exp_8}
(Color online)  Modeled (a) and measured (b) frequency response for experiment 3.}
\end{figure}

\Cref{fig:ray_model_data_freq_resp_comparision_exp_2,fig:ray_model_data_freq_resp_comparision_exp_8} compare modeled and measured evolution of the frequency responses for experiments 1 and 3 respectively. Modeled frequency response for experiment 1 captures many of the features seen in the measured response; increase in the frequency of modes; loss of consistent modal structure from ignition to \(T\) = 120~s; and the grouping of certain modes above 2000 Hz. The model for experiment 3 also captures many features of the experiment frequency response. While the models do not perfectly match the experimental results, very similar characteristics are seen. The models show differences between experiments 1 and 3 like that also present in the measured results. For example, in experiment 3 there is a complete loss of modal structure above 3500~Hz, which is not the case for experiment 1. The ray-traced models show this difference.

\begin{figure}[]
\centering
\includegraphics[angle=0,width=1.0\linewidth, keepaspectratio,,height=0.8\textheight]{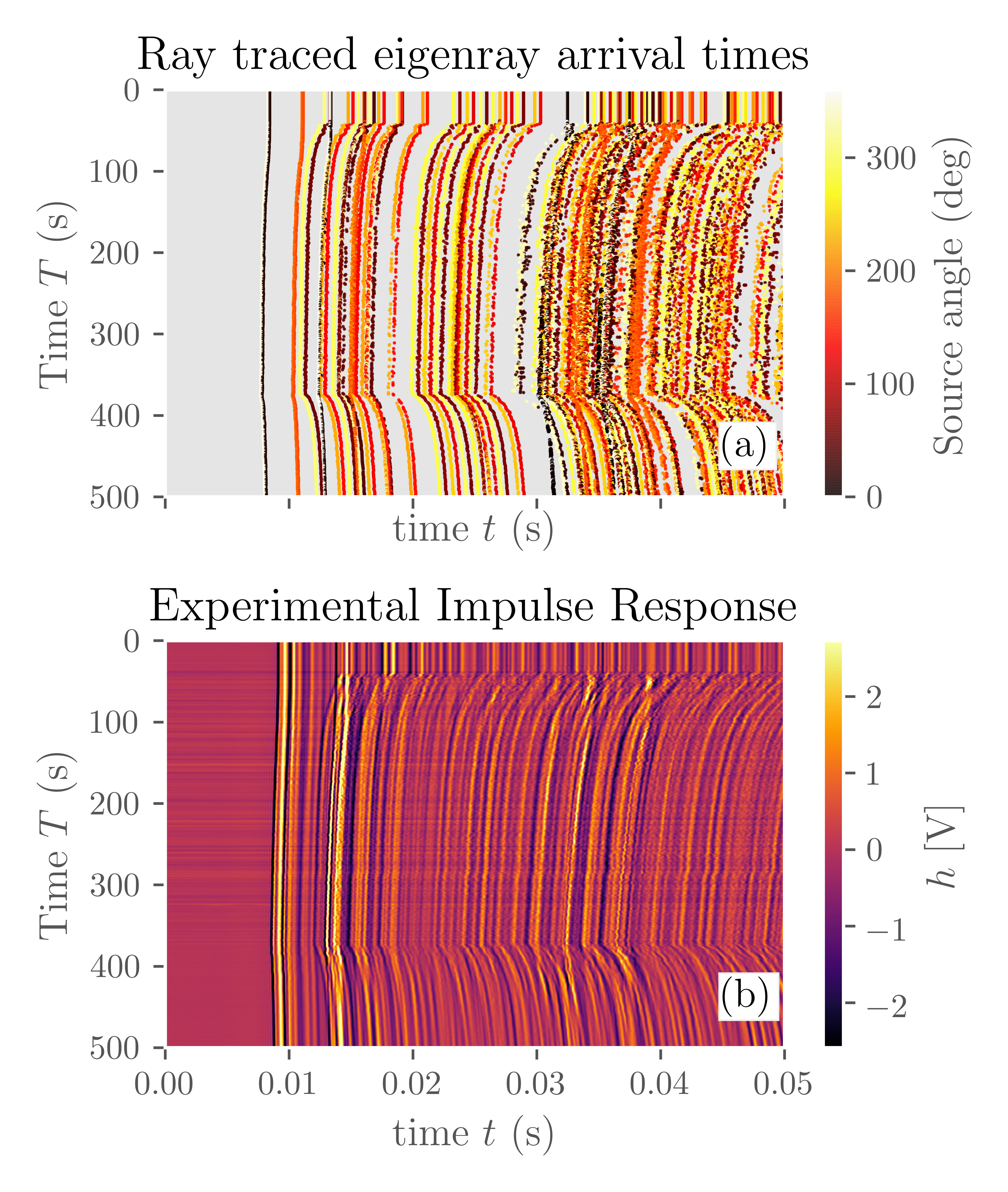}
\caption{\label{fig:ray_model_data_eigen_ray_comparision_exp_2}
(Color online)  Modeled (a) and measured (b) impulse response for experiment 1. The modeled response shows the arrival delay time, color coded by the source angle.}
\end{figure}

\begin{figure}[]
\centering
\includegraphics[angle=0,width=1.0\linewidth, keepaspectratio,,height=0.8\textheight]{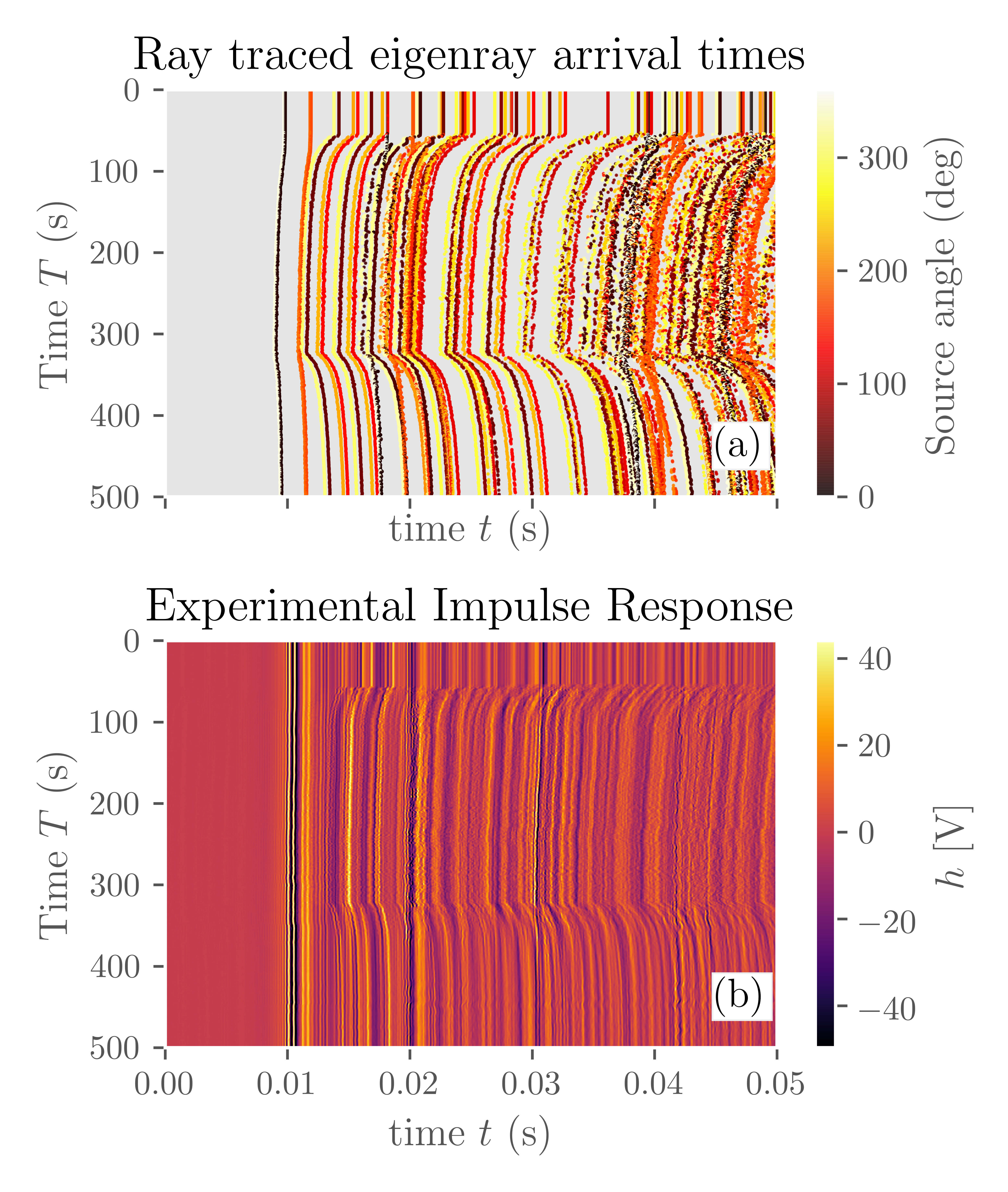}
\caption{\label{fig:ray_model_data_eigen_ray_comparision_exp_8}
(Color online)  Modeled (a) and measured (b) impulse response for experiment 3. The modeled response shows the arrival delay time, color coded by the source angle.}
\end{figure}

\Cref{fig:ray_model_data_eigen_ray_comparision_exp_2} and \Cref{fig:ray_model_data_eigen_ray_comparision_exp_8} compare modeled eigenray delay time, and experimentally measured impulse response for experiment 1 and 3 respectively. A distinction is made in this section between eigenray delays and the measured impulse responses. The eigenray delays are a perfect impulse response (i.e. the response of a discrete delta function, to the limit of float point 64-bit precision).  In contrast, the measured impulse responses are the response to bandwidth-limited finite duration pulse. This is an important distinction because the eigenray delays show much finer time resolution than the measured impulse responses. The eigenray delay time shows remarkably similar patterns as the measured impulse response, despite the model being two-dimensional and the experimental being three-dimensional. The earlier arrivals are the least impacted and more stable. Later arrivals have a larger change in delay time. In addition to the decreasing delay time after ignition, there is also a random spread in the times. The marker color in the eigenray plot indicates the elevation angle of the ray at the source (angle from the horizontal, positive is towards the ceiling). Observe that the shallow rays arrive at the receiver earlier (direct path and shallow bottom reflection paths) and are impacted the least after ignition. The early arrivals being least impacted by the fire is consistent with the measured data. Certain key features are captured very well. For example, both modeled and measured data show a crossing of arrival paths due to the fire, i.e. paths that arrived earlier in iso-speed arrive later than other paths after ignition. 

The ray models were run with identical parameters, except for the geometry and fire configuration. The differences between model results are qualitatively like the differences between the experimental results. While the comparisons between model and data are not perfect, the model captures many of the features seen in the data. This lends validity to the model, and to the assertion that the temperature variation is a major cause of the frequency response change. 

\section{Conclusion}
\label{sec:org6abf2e4}

Sound propagation in a compartment fire was modeled using two acoustic modeling modalities, ray theory, and full-wave finite element modeling. Both models relied on environmental inputs from a CFD fire model. Good agreement was found in measured and modeled frequency and impulse responses. The results show that temperature variations due to the fire can account for many of the observed phenomena.

Future work into this problem should explore three-dimensional modeling with frequency-dependent losses and the effect of horizontal temperature gradients. 

\section{Acknowledgement}
\label{sec:orgb853aa1}
The experimental measurements were funded by the U.S. Department of Homeland Security Assistance to Firefighters Grants Program. Analysis was self-funded by Dr. Abbasi. The authors thank Mudeer Habeeb, Kyle Ford, and Joelle Suits for their assistance with the experiments. 

\section{References}
\label{sec:orgf1c4971}

\bibliography{Chapter_1_master}

\end{document}